\documentclass[prb,preprint]{revtex4-1}

\usepackage{amsmath}  
\usepackage{amsfonts} 
\usepackage{graphicx} 
\usepackage{extarrows}
\usepackage[colorlinks,
            linkcolor=red,
            anchorcolor=green,
            citecolor=blue
            ]{hyperref}
\begin{document}

\title{Analytic solution to the motion of the mass-spring oscillator subjected to external force}


\author{Youtian Zhang}
\email{ytzhang@smail.nju.edu.cn} 
\affiliation{Department of Physics, Nanjing University, Nanjing 210093, China
\\Department of Electrophysics, National Chiao Tung University, Hsinchu, China}

\begin{abstract}
The non-evanescent vibration, evanescent vibration and forced vibration of an oscillator attached to the massless spring are always discussed in general mechanics courses. In this article, we focus on the heavy-spring conditions. We first investigate the general situation where both viscous resistance and applied force are considered under the perspective of the renormalization group theory. Then we deal with the special case where renormalization method fails to work. We use analytic method to study the evanescent vibration of an oscillator attached to the heavy spring. Generalized orthogonal and over-complete base set is established. Two independent vibration modes indicate the vibrational frequency becomes bigger than simple harmonic oscillator if the mass of spring is considered.
\end{abstract}

\maketitle 

\section{Introduction} 
An oscillation is a common but very important phenomenon in the physical world. If a physical quantity is displaced from the equilibrium a little, negative feedback may then lead to an oscillation. A familiar example is a simple harmonic oscillator. Also, evanescent vibrations and forced vibrations of an oscillator are normally focused.\cite{la} The mass of the spring is neglected in models. However, the mass of the spring is unnecessarily neglected. In this article we try to solve the mass-spring system where the mass of the spring is not negligible.

In 1979, Weinstock studied the normal modes of the oscillator motion for the oscillator attached to a heavy spring by virtue of the Stieltjes integral\cite{rw}. In 1994, Nunes da Silva obtained the normal frequencies of elastic oscillations of a particle suspended on a spring of non-negligible mass again under the perspective of the renormalization group theory\cite{re}.

A continuous spring can be regarded as a chain of many small springs coupling an equal amount of small masses. We then repeat mapping process by associating two consecutive small springs into a single one. At last, we only need to focus our attention on boundary effects. da Silva dealt with the problem only for the simplest situation, without the friction and applied force, and the advantage of this method does not emerge in the simplest case. In fact, we can not only find out the normal frequencies in the conservative system, but also obtain the specific equation of motion when external forces are acted on. In the next sections, we first explore the most general condition, a forced vibration with friction, with the help of the renormalization method. Then we deal with a special case analytically where the renormalization method can fail to work. We derive the equations and investigate the generalized orthogonality of the base set of the solution, and then obtain the normal modes of a damping oscillator attached to a heavy spring.
\section{The forced vibration with friction considering the mass of the spring}
Hang an uniformly distributed spring with mass $m$ vertically. The top side is fastened to stable fixture and the bottom side concatenate an object with mass $M$ as oscillator. The free length of spring is $L$ and the stiffness coefficient is $k$. To discretize the heavy spring, we view it as a series of $N$ equal and small non-mass springs and each small spring coupling a concentrated object with mass $m/N$. The natural length and elastic constant(labeled by $s$) of each small spring are $L/N$ and $s=kN$ respectively. The damping on the oscillator $M$ can be calculated as $-bv_M$ if the velocity of oscillator is $v_M$ and damping coefficient is $b$. Besides, we applied a time-dependent force $f(t)$ on the oscillator.
\begin{figure}[htbp]
\centering
\includegraphics[height=8cm]{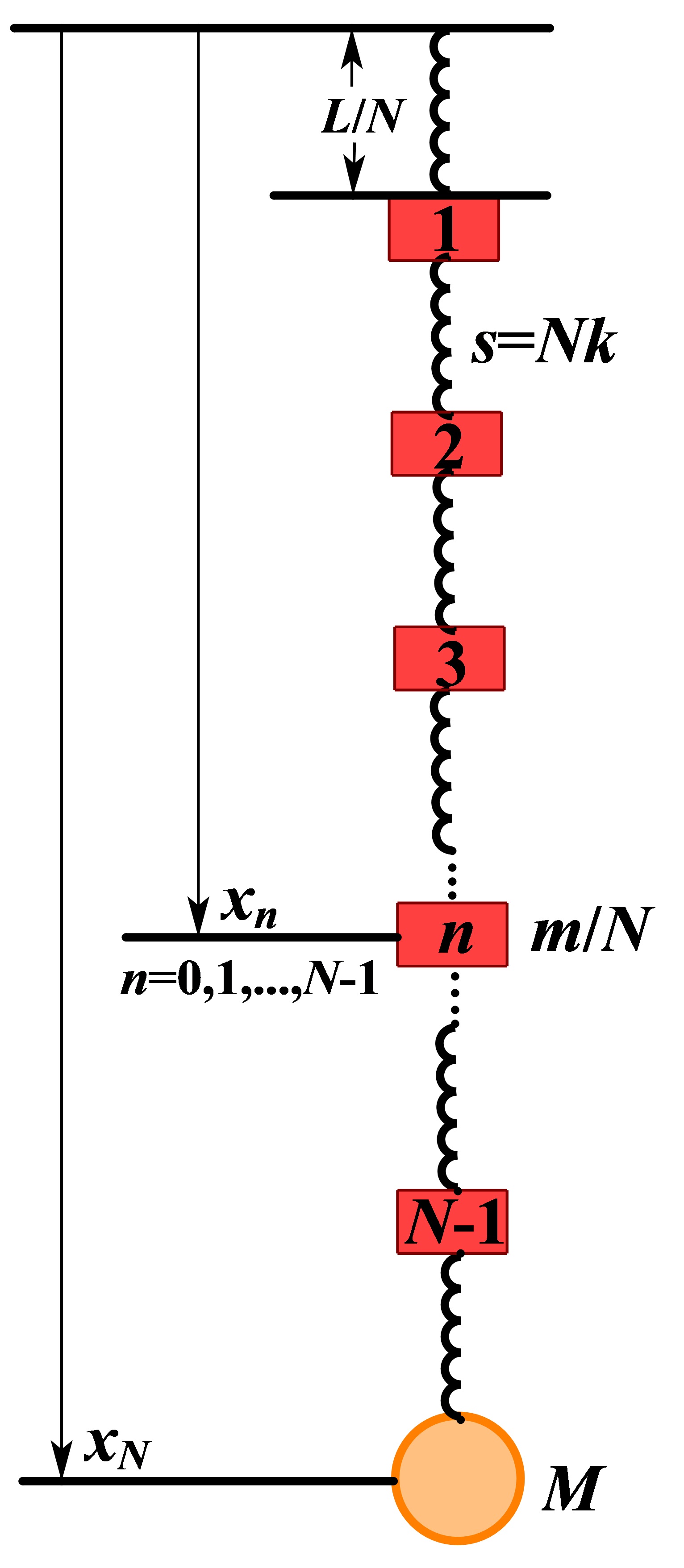}
\caption{A discretization model for the heavy spring with natural length $L$, mass $m$ and stiffness coefficient $k$ when divided into $N$ small springs coupling equal amount concentrated objects.}
\end{figure}

The positions of objects are denoted by $x_i(i=0,1,...,N)$, then equations of motion can be set up as
\begin{equation}
\begin{aligned}
&x_0=0,\\ \label{yi}
&\frac{m}{N}\frac{\textrm{d}^2}{\textrm{d}t^2}x_n=s(x_{n-1}-2x_n+x_{n+1})+\frac{m}{N} g\quad(1\leq n\leq N-1),\\
&M\frac{\textrm{d}^2}{\textrm{d}t^2}x_N=-s(x_N-x_{N-1}-\frac{L}{N})+Mg-b\frac{\textrm{d}}{\textrm{d}t}x_N+f(t).
\end{aligned}
\end{equation}
We can eliminate the constant terms derived from gravity in the Eq.\eqref{yi} by changing the coordinates appropriately. So that we can use newly defined coordinates($u_i,i=0,1,...,N$) to describe the motion. J.M.N. Dasilva proposed a way to approach these newly defined coordinates.\cite{re} The equations with new coordinates are written as follows,
\begin{equation}
\begin{aligned}
&u_0=0,\\  \label{er}
&\frac{m}{N}\frac{\textrm{d}^2}{\textrm{d}t^2}u_n=s(u_{n-1}-2u_n+u_{n+1})\quad(1\leq n\leq N-1),\\
&M\frac{\textrm{d}^2}{\textrm{d}t^2}u_N=-s(u_N-u_{N-1})-b\frac{\textrm{d}}{\textrm{d}t}u_N+f(t).
\end{aligned}
\end{equation}
If we denote $\tilde{f}=\frac{1}{s}\mathcal{F}[f(t)] =\frac{1}{Nk}\int\limits_{-\infty}^\infty f(t) e^{-i\omega t}\,\textrm{d}t$ and $U_n=\mathcal{F}[u_n(t)] = \int\limits_{-\infty}^\infty u_n(t) e^{-i\omega t}\,\textrm{d}t(n=0,1,...,N)$, then the Eq.\eqref{er} can be Fourier transformed as
\begin{equation}
\begin{aligned}
&U_0=0,\\ \label{fo}
&cU_n=U_{n-1}+U_{n+1}\quad(1\leq n\leq N-1),\\
&CU_N=U_{N-1}+\tilde{f},
\end{aligned}
\end{equation}
with $c=2(1-\frac{m\omega^2}{2N^2k})$ and $C=1+\frac{i\omega b-M\omega^2}{Nk}$.

By Eq.\eqref{fo}, $c^2U_{2n}=c(U_{2n-1}+U_{2n+1})=U_{2n-2}+2U_{2n}+U_{2n+2}$ and $cCU_N=cU_{N-1}+c\tilde{f}=U_{N-2}+U_N+c\tilde{f}$. We combine two small springs into a bigger small spring. That is, with this mapping process, the previous $(2n-1)^{\textrm{th}}$ and $(2n)^{\textrm{th}}$ springs now become the $n^{\textrm{th}}$ bigger small spring. The new Fourier transformed position function of $n^{\textrm{th}}$ spring is denoted by $\overline{U_n}$ and $\overline{U_n}=U_{2n}$. So we have
\begin{subequations}
\begin{align}
&(c^2-2)\overline{U_n}=\overline{U_{n-1}}+\overline{U_{n+1}}\label{s1}\\
&(cC-1)\overline{U_{\frac{N}{2}}}=\overline{U_{\frac{N}{2}-1}}+c\tilde{f}\label{s2}
\end{align}
\label{sa}
\end{subequations}
Comparing Eq.\eqref{sa} and Eq.\eqref{fo}, the equations change regularly\cite{6} after the process of combining two consecutive small springs into bigger spring. So we repeat the combination to renormalize. If we set $N=2^p$, then after $p^{\textrm{th}}$ repeat, Eq.\eqref{s2} finally becomes $C^{(p)}\overline{U_1}=\overline{U_0}+\tilde{f}\displaystyle\prod_{\kappa=0}^{p-1}c^{(\kappa)}$. $\overline{U_0}=0$ and the position function of the oscillator is $
u_{\textrm{oscillator}}(t)=\frac{1}{2\pi}\displaystyle \int\limits_{-\infty}^\infty \frac{\tilde{f}\displaystyle\prod_{\kappa=0}^{p-1}c^{(\kappa)}}{C^{(p)}} e^{i\omega t}\,\textrm{d}\omega$.

The next thing to do is to find out the iteration value $C^{(p)}$ and $\displaystyle\prod_{\kappa=0}^{p-1}c^{(\kappa)}$. Compare between Eqs.\eqref{fo} and \eqref{sa} gives $c^{(1)}=c^2-2$ and $C^{(1)}=cC-1$, so $2C^{(1)}-c^{(1)}=c(2C-c)$ and more generally,
\[
2C^{(p)}-c^{(p)}=(2C-c)\displaystyle\prod_{\kappa=0}^{p-1}c^{(\kappa)}.
\]
Introduce $\gamma$ and let $c\equiv2\textrm{cos}\gamma$. $c^{(p)}$ and $\displaystyle\prod_{\kappa=0}^{p-1}c^{(\kappa)}$ are easily accessible with this variable substitution which are $c^{(p)}=2\textrm{cos}2^{p}\gamma$ and $\displaystyle\prod_{\kappa=0}^{p-1}c^{(\kappa)}=\frac{\textrm{sin}2^{p}\gamma}{\textrm{sin}\gamma}$. Finally,
\[
C^{(p)}=\frac{1}{2}[\frac{\textrm{sin}2^{p}\gamma}{\textrm{sin}\gamma}(2C-c)+2\textrm{cos}2^{p}\gamma].
\]
$c\equiv2\textrm{cos}\gamma=2(1-\frac{m\omega^2}{2N^2k})$. Notice that $N\gg1$, so $\gamma\ll1$ and $\gamma=\textrm{sin}\gamma=\frac{\omega}{N}\sqrt{\frac{m}{k}}.$
\begin{equation}
\begin{aligned}
u_{\textrm{oscillator}}(t)&=\frac{1}{2\pi} \int\limits_{-\infty}^\infty \frac{\tilde{f}\displaystyle\prod_{\kappa=0}^{p-1}c^{(\kappa)}}{C^{(p)}} e^{i\omega t}\,\textrm{d}\omega\\
&=\frac{1}{2\pi} \int\limits_{-\infty}^\infty \frac{2\textrm{sin}(\omega\sqrt{\frac{m}{k}})\tilde{f}}{\textrm{sin}(2^{p}\gamma)(2C-c)+2\textrm{sin}\gamma \textrm{cos}(2^{p}\gamma)} e^{i\omega t}\,\textrm{d}\omega\\
&=\frac{1}{2\pi}\int\limits_{-\infty}^\infty \frac{\int\limits_{-\infty}^\infty f(t) e^{-i\omega t}\,\textrm{d}t}{i\omega b-M\omega^2+\textrm{cot}(\omega\sqrt{\frac{m}{k}})\omega\sqrt{km}}e^{i\omega t}\,\textrm{d}\omega.
\end{aligned}
\end{equation}
It's the motion equation of the oscillator.\cite{7} But one thing to note here is that $f(t)$ can not be $0$ or the solution vanishes. We deal with this condition in following part.

In fact, if $g(t)\equiv\frac{1}{2\pi}{\int\limits_{-\infty}^\infty \frac{1}{-i\omega b-M\omega^2+\textrm{cot}(\omega\sqrt{\frac{m}{k}})\omega\sqrt{km}} e^{i\omega t}\,\textrm{d}\omega}$, the motion equation of the oscillator can be given by the convolution of $f(t)$ and $g(t)$, i.e. $u_{\textrm{oscillator}}(t)=f(t)\ast g(t)$. From the following discussion, we will know that the frequency which satisfies $-i\omega b-M\omega^2+\textrm{cot}(\omega\sqrt{\frac{m}{k}})\omega\sqrt{km}=0$ is exactly the eigenvalue in no applied force condition.
\section{No applied force condition}
Since we can no longer use renormalization method for no applied force condition, we then use mathematical physics equations to study this problem. Adopt appropriate coordinates as introduced before, and the problem can be analytically described as follows.\cite{8}
\begin{equation}
\begin{cases}
u_{tt}-\frac{kL^2}{m}u_{xx}=0  \quad(t>0,x\in[0,L))\\ u(0,t)=0\\u_x(L,t)=-\frac{M}{kL}u_{tt}(L,t)-\frac{b}{kL}u_t(L,t)\\
\left.u\right|_{t=0}=\phi(x)\\
\left.\frac{\partial u}{\partial t}\right|_{t=0}=\psi(x)\end{cases}\label{eq}
\end{equation}
We consider using method of separation of variables to solve this equation and we take $u(x,t)=X(x)\textrm{exp}(-i\mu L\sqrt{\frac{k}{m}}t)$ as the ansatz. The equations above then become
\begin{equation}\label{0}
\begin{cases}
X''(x)+\mu^2X(x)=0, \\ X(0)=0,\\X'(L)=(i\mu\frac{b}{\sqrt{km}}+\mu^2\frac{ML}{m})X(L),\\X(x)=\phi(x),\\
-i\mu L\sqrt{\frac{k}{m}}X(x)=\psi(x).\end{cases}
\end{equation}
To investigate the base set of solutions, we set $X_p(x)$ and $X_q(x)$ are solutions to the equations above from the base set.
\begin{equation}
X_p''(x)+\mu_p^2X_p(x)=0\label{fir}
\end{equation}
\begin{equation}
X_p'(L)=(i\mu_p\frac{b}{\sqrt{km}}+\mu_p^2\frac{ML}{m})X_p(L)\label{sec}
\end{equation}
\begin{equation}
X_q''(x)+\mu_q^2X_q(x)=0\label{thi}
\end{equation}
\begin{equation}
X_q'(L)=(i\mu_q\frac{b}{\sqrt{km}}+\mu_q^2\frac{ML}{m})X_q(L)\label{fou}
\end{equation}
\eqref{sec}$\times X_q(L)-$\eqref{fou}$\times X_p(L)$ and we get
\begin{equation}
\begin{aligned}
&\frac{ML}{m}(\mu_p^2-\mu_q^2)X_p(L)X_q(L)+i(\mu_p-\mu_q)\frac{b}{\sqrt{km}}X_p(L)X_q(L)\\
=&X_q(L)X'_p(L)-X'_q(L)X_p(L)\\
=&\left.X_q(x)X'_p(x)\right|_0^L-\int_0^LX'_p(x)X'_q(x)\textrm{d}x-\left.X_p(x)X'_q(x)\right|_0^L+\int_0^LX'_p(x)X'_q(x)\textrm{d}x\\
=&\int_0^LX_q(x)d[X'_p(x)]-\int_0^LX_p(x)d[X'_q(x)]\\
=&\int_0^LX_q(x)X''_p(x)\textrm{d}x-\int_0^LX_p(x)X''_q(x)\textrm{d}x\\
=&-\mu_p^2\int_0^LX_q(x)X_p(x)\textrm{d}x+\mu_q^2\int_0^LX_p(x)X_q(x)\textrm{d}x.
\end{aligned}\label{5}
\end{equation}
From eq.\eqref{5} we eventually reach the following equality,
\begin{equation}
\int_0^LX_q(x)X_p(x)dx+\left[\frac{ML}{m}+i\frac{b}{(\mu_p+\mu_q)\sqrt{km}}\right]X_p(L)X_q(L)=0.\label{6}
\end{equation}
By definition, eq.\eqref{6} shows that the solutions in the base set are generalized orthogonal\cite{pde}. The square modulus (denoted by $N^2$) of the eigenfunctions can be calculated as
\begin{equation}\label{n2}
\begin{aligned}
N^2[X_p(x)]&=\int_0^LX_p^2(x)dx+\left[\frac{ML}{m}+i\frac{b}{2\mu_p\sqrt{km}}\right]X_p^2(L)\\
&\xlongequal[\textrm{Eq}.\eqref{sec}]{X_p(x)=\textrm{sin}\mu_px}\frac{L}{2}-\frac{1}{4\mu_p}\textrm{sin}2\mu_pL+\frac{\textrm{cos}\mu_pL\textrm{sin}\mu_pL}{2\mu_p}+\frac{ML}{2m}\textrm{sin}^2\mu_pL\\
&=\frac{L}{2}+\frac{ML}{2m}\textrm{sin}^2\mu_pL.
\end{aligned}
\end{equation}
So the solution to the Eq.\eqref{eq} can be written as $u(x,t)=\sum_nA_n\textrm{sin}\mu_nx \textrm{exp}(-i\mu_nL\sqrt{\frac{k}{m}}t)$ and the expansion coefficient $A_n$ can be determined from the initial conditions. $\phi(x)$ and $\psi(x)$ could be expanded as
$\phi(x)=\sum_nP_n\textrm{sin}\mu_nx$, $\psi(x)=-i\sum_n\mu_n L\sqrt{\frac{k}{m}}Q_n\textrm{sin}\mu_nx$ where
\begin{equation}
\begin{aligned}\label{A}
P_n=&\frac{\int_0^L\phi(x)\textrm{sin}\mu_nxdx+\left[\frac{ML}{m}+i\frac{b}{2\mu_n\sqrt{km}}\right]\phi(L)\textrm{sin}\mu_nL}{N^2[X_n(x)]},\\
Q_n=&\frac{\int_0^L\frac{i}{\mu_nL}\sqrt{\frac{m}{k}}\psi(x)\textrm{sin}\mu_nxdx+\left[\frac{ML}{m}+i\frac{b}{2\mu_n\sqrt{km}}\right]\frac{i}{\mu_nL}\sqrt{\frac{m}{k}}\psi(L)\textrm{sin}\mu_nL}{N^2[X_n(x)]}.
\end{aligned}
\end{equation}
Let $A_n=\alpha P_n+\beta Q_n$, we then have
\begin{equation}
\begin{aligned}\label{ab}
\phi(x)&=\sum_n(\alpha P_n+\beta Q_n)\textrm{sin}\mu_nx,\\
\psi(x)&=-i\sum_n\mu_n L\sqrt{\frac{k}{m}}(\alpha P_n+\beta Q_n)\textrm{sin}\mu_nx.
\end{aligned}
\end{equation}
Compare Eq.\eqref{ab} with $\phi(x)=\sum_nP_n\textrm{sin}\mu_nx$, $\psi(x)=-i\sum_n\mu_n L\sqrt{\frac{k}{m}}Q_n\textrm{sin}\mu_nx$,
\begin{equation}
\begin{aligned}
P_n[\alpha(1-i\mu_nL\sqrt{\frac{k}{m}})-1]+Q_n[\beta(1-i\mu_nL\sqrt{\frac{k}{m}})+i\mu_nL\sqrt{\frac{k}{m}}]=0,\\
P_n[\alpha(1+i\mu_nL\sqrt{\frac{k}{m}})-1]+Q_n[\beta(1+i\mu_nL\sqrt{\frac{k}{m}})-i\mu_nL\sqrt{\frac{k}{m}}]=0.
\end{aligned}
\end{equation}
Hence, $\alpha=\frac{1}{1\pm i\mu_n L\sqrt{\frac{k}{m}}}$ and $\beta=\frac{\pm i \mu_nL\sqrt{\frac{k}{m}}}{1\pm i\mu_nL\sqrt{\frac{k}{m}}}$. Finally $A_n$ can be given as
\begin{equation}\label{an}
A_n=\frac{1}{N^2[X_n(x)](1\pm i\mu_nL\sqrt{\frac{k}{m}})}\left[\int_0^L[\phi(x)\mp\psi(x)]\textrm{sin}\mu_nx\textrm{d}x+[\frac{ML}{m}+i\frac{b}{2\mu_n\sqrt{km}}][\phi(L)\mp\psi(L)]\textrm{sin}\mu_nL\right].
\end{equation}
This non-unique expansion also suggests that the generalized orthogonality of base set $\{X_n(x)\}$ is over-complete.

\begin{figure}
\centering
\begin{minipage}[t]{0.48\linewidth}
    \centering
    \includegraphics[height=4.5cm]{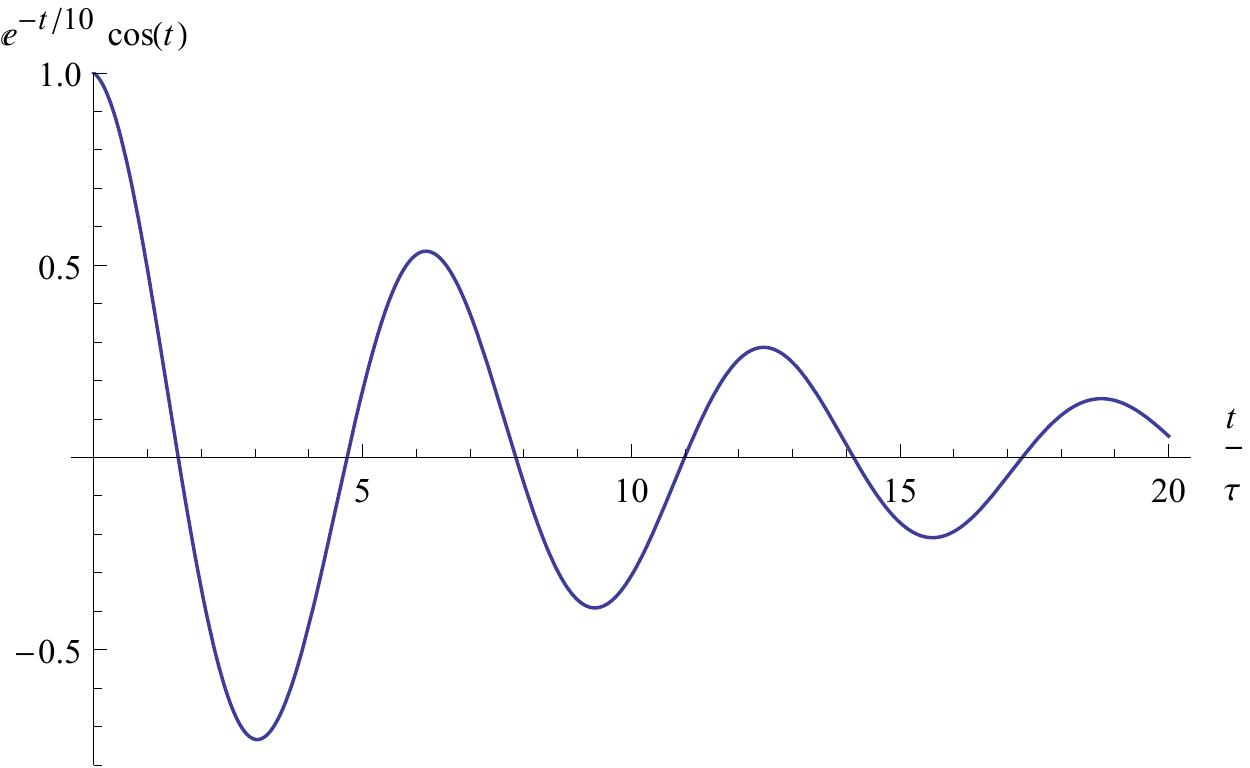}
\end{minipage}
\hspace{3ex}
\begin{minipage}[t]{0.45\linewidth}
    \centering
    \includegraphics[height=4.5cm]{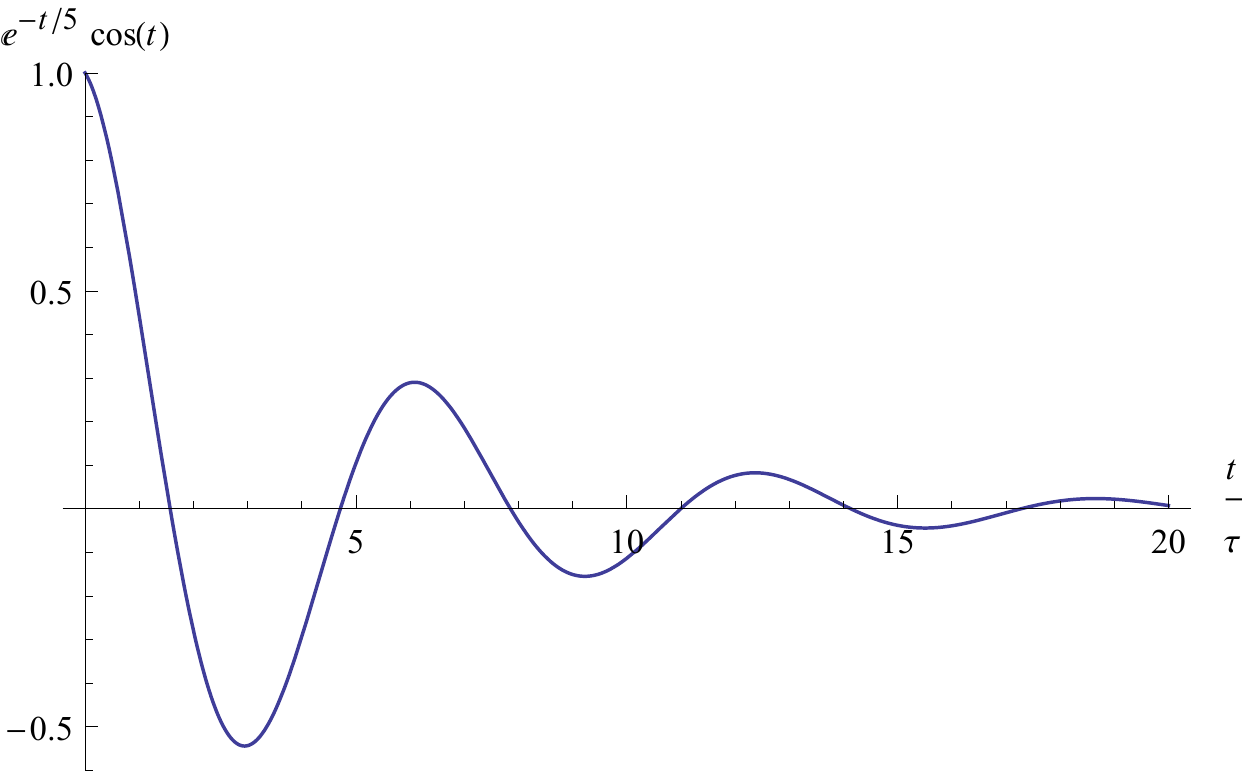}
\end{minipage}\\
\caption{Let $t$ be dimensionless with $\tau=\sqrt{\frac{m}{k}}/\xi_nL$, the eigen-vibration mode of the oscillator can be plotted under two cases with $\zeta_n/\xi_n=0.1$ and $\zeta_n/\xi_n=0.2$. The oscillator vibrates with damped amplitude.}\label{res}
\end{figure}

The eigenvalue equation reveals as $\textrm{cot}\mu L=i\frac{b}{\sqrt{km}}+\mu\frac{ML}{m}$ (eigenvalue $\mu$ won't be 0). Notice that $-\overline{\mu}$ is also eigenvalue if $\mu$ is eigenvalue. The corresponding expansion coefficients have the relation $P(-\overline{\mu})=-\overline{P(\mu)}$ and $Q(-\overline{\mu})=-\overline{Q(\mu)}$ due to the Eq.\eqref{A}.
Considering $\textrm{sin}(-\overline{\mu}x)=-\overline{\textrm{sin}\mu x}$,
\begin{equation}
\begin{aligned}\label{ev}
&\sum_nP_n\textrm{sin}\mu_nx=\displaystyle{\sum_{\textrm{Re}(\mu_n)>0}}[P_n\textrm{sin}\mu_nx+\overline{P_n}\textrm{sin}\overline{\mu_n}x]=\displaystyle{\sum_{\textrm{Re}(\mu_n)>0}}[P_n\textrm{sin}\mu_nx+c.c.],\\
&-i\sum_n\mu_n L\sqrt{\frac{k}{m}}Q_n\textrm{sin}\mu_nx=-i[\displaystyle{\sum_{\textrm{Re}(\mu_n)>0}}[\mu_n L\sqrt{\frac{k}{m}}Q_n\textrm{sin}\mu_nx]-c.c.].
\end{aligned}
\end{equation}
Eq.\eqref{ev} verifies that both $\phi(x)$ and $\psi(x)$ are pure real when ansatz $u(x,t)=\sum_nA_n\textrm{sin}\mu_nx \textrm{exp}(-i\mu_nL\sqrt{\frac{k}{m}}t)$ is taken. Of course $u(x,t)=\displaystyle{\sum_{\textrm{Re}(\mu_n)>0}}[A_n\textrm{sin}\mu_nx\textrm{exp}(-i\mu_nL\sqrt{\frac{k}{m}}t)+c.c.]$ is also pure real. We can obtain two independent eigen-vibration modes from this result. Let $\mu_n=\xi_n-i\zeta_n$\cite{9}, then
\begin{equation}
\begin{aligned}
&\quad\textrm{sin}\mu_nx\textrm{exp}(-i\mu_nL\sqrt{\frac{k}{m}}t)\\
&=\textrm{sin}(\xi_n-i\zeta_n)x\textrm{exp}[-i(\xi_n-i\zeta_n)L\sqrt{\frac{k}{m}}t]\\
&=(\textrm{sin}\xi_nx\textrm{cosh}\zeta_nx-i\textrm{cos}\xi_nx\textrm{sinh}\zeta_nx)\textrm{exp}(-\zeta_nL\sqrt{\frac{k}{m}}t)[\textrm{cos}(\xi_nL\sqrt{\frac{k}{m}}t)-i\textrm{sin}(\xi_nL\sqrt{\frac{k}{m}}t)]\\
&=[\textrm{sin}\xi_nx\textrm{cosh}\zeta_nx\textrm{cos}(\xi_nL\sqrt{\frac{k}{m}}t)-\textrm{cos}\xi_nx\textrm{sinh}\zeta_nx\textrm{sin}(\xi_nL\sqrt{\frac{k}{m}}t)]\textrm{exp}(-\zeta_nL\sqrt{\frac{k}{m}}t)-\\
&\quad i[\textrm{cos}\xi_nx\textrm{sinh}\zeta_nx\textrm{cos}(\xi_nL\sqrt{\frac{k}{m}}t)+\textrm{sin}\xi_nx\textrm{cosh}\zeta_nx\textrm{sin}(\xi_nL\sqrt{\frac{k}{m}}t)]\textrm{exp}(-\zeta_nL\sqrt{\frac{k}{m}}t).
\end{aligned}
\end{equation}
The two independent vibration modes are given by
\[
\begin{aligned}
\textrm{mode 1:}[\textrm{sin}\xi_nx\textrm{cosh}\zeta_nx\textrm{cos}(\xi_nL\sqrt{\frac{k}{m}}t)-\textrm{cos}\xi_nx\textrm{sinh}\zeta_nx\textrm{sin}(\xi_nL\sqrt{\frac{k}{m}}t)]\textrm{exp}(-\zeta_nL\sqrt{\frac{k}{m}}t),\\
\textrm{mode 2:}[\textrm{cos}\xi_nx\textrm{sinh}\zeta_nx\textrm{cos}(\xi_nL\sqrt{\frac{k}{m}}t)+\textrm{sin}\xi_nx\textrm{cosh}\zeta_nx\textrm{sin}(\xi_nL\sqrt{\frac{k}{m}}t)]\textrm{exp}(-\zeta_nL\sqrt{\frac{k}{m}}t).
\end{aligned}
\]
The oscillator ($x=L$) vibrates with damped amplitude (as is shown in Fig.\ref{res}) in both modes, which is reasonable.

For zero-friction case ($b=0$), the solutions to eigenvalue equation $\textrm{cot}\mu L=i\frac{b}{\sqrt{km}}+\mu\frac{ML}{m}$ are pure real, i.e. $\zeta=0$. Then two independent vibration modes become
\[
\begin{aligned}
\textrm{mode 1:}\quad\textrm{sin}\xi_nx\textrm{cos}(\xi_nL\sqrt{\frac{k}{m}}t),\\
\textrm{mode 2:}\quad\textrm{sin}\xi_nx\textrm{sin}(\xi_nL\sqrt{\frac{k}{m}}t).
\end{aligned}
\]
With $\textrm{cot}\xi_n L=\frac{M}{m}\xi_n L$, we can also obtain $\xi_nL\sqrt{\frac{k}{m}}>\sqrt{\frac{k}{M}}$. So the vibrational frequency becomes bigger than simple harmonic oscillator if the mass of spring is considered.\cite{10}

Summarize the result, the solution to Eq.\eqref{eq} is $u(x,t)=\sum_nA_n\textrm{sin}\mu_nx \textrm{exp}(-i\mu_nL\sqrt{\frac{k}{m}}t)$ where eigenvalue $\mu_n$ is given by $\textrm{cot}\mu_n L=i\frac{b}{\sqrt{km}}+\mu_n\frac{ML}{m}$. Expansion coefficient $A_n$ and modulus square are given by Eqs.\eqref{an} and \eqref{n2}.
\section{Conclusion}
In this article, we detailedly studied the vibration of spring oscillator when the mass of the spring can't be neglect. Damped oscillation and forced vibration are especially focused. For general condition, oscillation with friction and applied force, renormalization method is employed to obtain the equation of the motion. Renormalization method shows superiority when there is applied force $f(t)$ exerts on the oscillator. We also investigate the damping vibration without applied force with theory of partial differential equations. For this specific boundary condition, the generalized orthogonality of base set is studied. The motion equation of this condition is also given. We discussed the characters of the eigenvalue and the expansion coefficient and the discussion verifies the validity of the solution.
\\\\\newline




\begin{thebibliography}{99}
\bibitem{7}
Notice that $\int\limits_{-\infty}^\infty \frac{k\int\limits_{-\infty}^\infty f(t) e^{-i\omega t}\,\textrm{d}t}{i\omega b-M\omega^2+\textrm{cot}(\omega\sqrt{\frac{m}{k}})\omega\sqrt{km}}e^{i\omega t}\,\textrm{d}\omega=\int\limits_{\infty}^{-\infty} \frac{k\int\limits_{-\infty}^\infty f(t) e^{-i(-\omega) t}\,\textrm{d}t}{i(-\omega) b-M(-\omega)^2+\textrm{cot}((-\omega)\sqrt{\frac{m}{k}})(-\omega)\sqrt{km}}e^{i(-\omega) t}\,\textrm{d}(-\omega)$, so $\int\limits_{-\infty}^\infty \frac{k\int\limits_{-\infty}^\infty f(t) e^{-i\omega t}\,\textrm{d}t}{i\omega b-M\omega^2+\textrm{cot}(\omega\sqrt{\frac{m}{k}})\omega\sqrt{km}}e^{i\omega t}\,\textrm{d}\omega=\int\limits_{-\infty}^\infty \frac{k\int\limits_{-\infty}^\infty f(t) e^{i\omega t}\,\textrm{d}t}{-i\omega b-M\omega^2+\textrm{cot}(\omega\sqrt{\frac{m}{k}})\omega\sqrt{km}}e^{-i\omega t}\,\textrm{d}\omega$ and $u_{\textrm{oscillator}}(t)$ is pure real.
\bibitem{8}
Take infinitesimal spring with length $\textrm{d} x$, stiffness coefficient $\frac{kL}{\textrm{d} x}$ and mass $\frac{\textrm{d} x}{L}m$. Suppose its amount of deformation is $\textrm{d} u$. Then its equation of motion in infinitesimal time $\textrm{d} t$ can be given by $\frac{\partial(\frac{kL}{\textrm{d} x}\textrm{d}u)}{\partial x}\textrm{d} x=\frac{\textrm{d} x}{L}m\cdot\frac{\textrm{d}^2u}{\mathrm{d} t^2}$ according to Newton's second law.
\bibitem{9}
It's hard to find all $\mu$s analytically from $\textrm{cot}\mu L=i\frac{b}{\sqrt{km}}+\mu\frac{ML}{m}$ but we can use perturbation to investigate since imaginary part in RHS of eigenvalue equation $\frac{b}{\sqrt{km}}$ is feeble in generally underdamping conditions. The zero-order approximation, $\mu^{(0)}\textrm{tan}\mu^{(0)} L=\frac{m}{ML}$ for zero friction situation ($b=0$), is discussed in previous work.\cite{rw} Obviously, solutions are real numbers and come in pairs ($-\mu^{(0)}$ is also solution if $\mu^{(0)}$ satisfies zero-order eigenvalue equation). Perturbation calculation shows the first order modification is negative pure imaginary number and the paired zero-order eigenvalues $\pm\mu^{(0)}$ share the same first order modification. It coincides with the previous analysis that $-\overline{\mu}$ is eigenvalue if $\mu$ is eigenvalue.
\bibitem{10}
If we neglect the mass of the spring, the frequency $\xi_nL\sqrt{\frac{k}{m}}$ degrades, $\lim\limits_{m\rightarrow0}\xi_nL\sqrt{\frac{k}{m}}=\sqrt{\frac{k}{M}}$, which can be proved by using $\lim\limits_{x\rightarrow0}\textrm{cot}x-\frac{1}{x}=0$.
\end{thebibliography}
\end{document}